# УПРАВЛЕНИЕ КОГЕРЕНТНОСТЬЮ ХАОТИЧЕСКИХ ОСЦИЛЛЯТОРОВ ПОСРЕДСТВОМ ДВУХ ЗАПАЗДЫВАЮЩИХ ОБРАТНЫХ СВЯЗЕЙ

*Голдобин Д.С., ст. преп. физического факультета ПГУ*
*Шкляева Е.В., ст. преп. математического факультета ПГУ*

Когерентность автоколебательных систем (иными словами, степень постоянности мгновенной частоты) является существенной их характеристикой. Она определяет качество часов, электрогенераторов, лазеров и т.п., а также предрасположенность колебательных систем к синхронизации и, следовательно, их восприимчивость к управляющему воздействию. Последнее также актуально и для хаотических систем, когда в них можно ввести фазоподобную переменную. Количественно когерентность может быть охарактеризована коэффициентом диффузии фазы колебаний:

$$D \equiv \lim_{t\to\infty} \left\langle (\varphi(t) - \varphi(0) - \langle\dot\varphi\rangle t)^2 \right\rangle / t \,. \qquad (1)$$

Использование слабой запаздывающей обратной связи позволяет изменять коэффициент диффузии фазы на порядки, не меняя тонкой структуры хаоса в системе [1]. Наиболее эффективного подавления диффузии удается добиться при временах запаздывания кратных среднему времени возврата Пуанкаре. Использование двух запаздывающих обратных связей с разными временами запаздывания предоставляет более гибкие и эффективные средства влияния на характеристики системы, чем единственная обратная связь [2,3]. В противоположность [4], мы интересуемся ситуациями, когда времена запаздываний несоизмеримы. Кроме того, некоторые системы обладают неотъемлемыми внутренними обратными связями с запаздыванием, и для них введение второй петли обратной связи является естественным и простым способом вмешательства с целью регулирования динамики. Эффективность использования второй петли обратной связи в таких системах обеспечивается схожестью механизмов влияния на систему для обеих обратных связей и, как следствие, возможностью их взаимной компенсации в широком диапазоне параметров. Например, в [5,6] предлагается использование дополнительной обратной связи для подавления нежелательной неустойчивости автомодуляции в лампе обратной волны.

В настоящей работе рассматривается применение двух запаздывающих обратных связей к хаотическим системам на примере системы Лоренца. Мы показываем, что уровни подавления или усиления коэффициента диффузии фазы, которых можно достичь, не разрушая хаос, при использовании обратных связей с двумя временами запаздывания существенно больше, чем при единственном времени запаздывания.

**Коэффициент диффузии фазы.** Конкретным рассматриваемым примером выбрана система уравнений Лоренца

$$\dot{x} = \sigma(y - x),$$
$$\dot{y} = rx - y - xz, \qquad (2)$$
$$\dot{z} = -bz + xy + k_1[z(t-\tau_1) - z(t)] + k_2[z(t-\tau_2) - z(t)],$$

при $\sigma = 10$, $r = 32$, $b = 8/3$ с запаздыванием в переменной $z$; $k_i$ и $\tau_i$ – коэффициенты и времена запаздывания обратных связей, соответственно. Фаза данной системы хорошо определена, если использовать проекцию траектории в плоскость $(u = \sqrt{x^2 + y^2}, z)$.

Результаты вычисления коэффициента диффузии $D$ представлены на Рис. 1. Можно видеть, что коэффициент диффузии особенно сильно чувствителен к обратным связям при $k_1 = k_2$ (Рис. 1a) на диагонали $\tau_1 = \tau_2$, что фактически соответствует случаю единственной запаздывающей связи с коэффициентом $k = 2k_{1,2}$. Однако, оказывается, что при совпадающих положительных коэффициентах связи $k_1 = k_2 > 0$ стабилизация нехаотических режимов и следующий за ней кризис хаоса происходят существенно раньше, чем при связях разного знака $k_1 = -k_2$ (Рис. 1b) или при $k_1 = k_2 < 0$ (Рис. 1c). На правом графике Рис. 1a в темных областях стационарные решения становятся устойчивыми (анализ устойчивости нехаотических режимов обсуждается ниже, в разделе "Мультистабильность"). Хотя при $|k_{1,2}| = 0.15$ хаотическое множество существует и остается устойчивым при всех рассмотренных $\tau_{1,2}$ (показатель Ляпунова нигде не опускается ниже $0.7$), появление сосуществующих устойчивых стационарных решений, во-первых, нежелательно само по себе, поскольку режим системы становится чувствителен к начальным условиям, теряется глобальная устойчивость хаотического аттрактора, во-вторых, свидетельствует о близости кризиса хаоса. Вместе с тем, минимальное значение коэффициента диффузии при $k_1 = k_2 = 0.15$: $D_{\min} \approx 0.0052$ достигается при $\tau_1 = \tau_2 \approx 0.56$;
при $k_1 = -k_2 = 0.15$: $D_{\min} \approx 0.0050$ – при $\tau_1 \approx 0.64$, $\tau_2 \approx 0.28$.

Таким образом, использование двух связей разного знака позволяет достичь того же уровня подавления диффузии фазы, не приближаясь к областям кризиса хаотического аттрактора.

**Мультистабильность.** В общем случае, достаточно сильная обратная связь может приводить к появлению локально устойчивых периодических траекторий [7] или стабилизации стационарных состояний [2,3]. Однако, известно, что метод Пирагаса (предполагающий использование единствен-ного времени запаздывания или цепочки кратных времен) не позволяет стабилизировать периодические траектории в системе Лоренца. Не наблюдалась такая стабилизация и в проводившемся в рамках настоящей работы численном счете с двумя временами запаздывания. Стабилизация стационарных состояний системы Лоренца при этом не исключена и именно она представляет опасность. Инкременты $\lambda$ возмущений этих состояний определяются следующим характеристическим уравнением:

$$\lambda(\lambda-\sigma-1)(\lambda+b+k_1(1-e^{-\lambda\tau_1})+k_2(1-e^{-\lambda\tau_2}))+b(r-1)(\lambda-2\sigma)=0. \quad (3)$$

Данное уравнение является трансцендентным, и результаты его численного решения – зависимость вещественной части максимального инкремента от времен запаздывания $\tau_i$ – представлены в правом столбце на Рис. 1.

**Заключение.** Исследовано влияние двух запаздывающих обратных связей на когерентность хаотических колебаний, количественно определяемую диффузией фазы, и выяснено, что коэффициент диффузии может меняться на 2–3 порядка. Совместный анализ хаотических свойств системы и устойчивости нехаотических режимов (стационарных состояний) выявляет целесообразность использования двух запаздывающих обратных связей, поскольку (например, при разных знаках коэффициентов связи, $k_1=-k_2$) это позволяет добиться на порядок большего подавления (или, при необходимости, усиления) диффузии без разрушения хаоса, чем единственная связь [1].

## Библиографический список

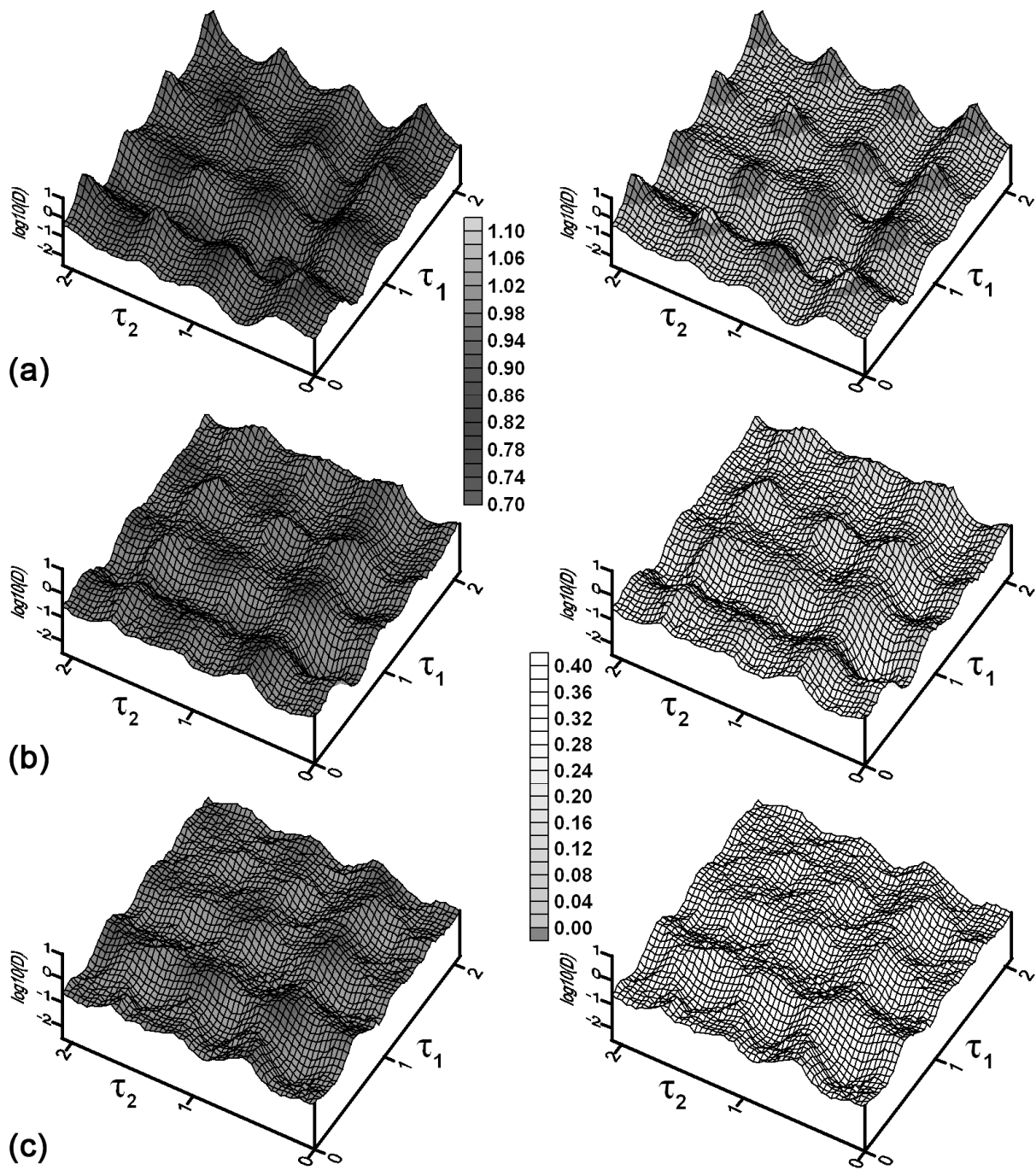

Рис. 1. Коэффициент диффузии фазы в системе Лоренца при двух запаздывающих обратных связях: (a) $k_1 = k_2 = 0.15$, (b) $k_1 = -k_2 = 0.15$, (c) $k_1 = k_2 = -0.15$.
Графики слева и справа отличаются только цветом: в левом столбце цвет представляет ведущий показатель Ляпунова для хаотического аттрактора, в правом – для нетривиальных стационарных решений.